\documentclass[a4paper,11pt]{article}
\pdfoutput=1 
\usepackage{jheppub} 
\newcommand{\Rdr}{R_{\Delta R}}
\newcommand{\Rdphi}{R_{\Delta \phi}}
\newcommand{\Rtt}{R_{3/2}}

\newcommand{\Dphimax}{\Delta \phi_{\rm max}}

\newcommand{\DR}{\Delta R}

\newcommand{\ptmax}{p_{T}^{\rm max}}
\newcommand{\ptmin}{p_{T \rm min}}
\newcommand{\ptngbmin}{p_{T \rm min}^{\rm nbr}}

\newcommand{\mur}{\mu_r}
\newcommand{\muf}{\mu_f}
\newcommand{\murf}{\mu_{r,f}}
\newcommand{\as}{\alpha_s}
\newcommand{\ass}{\alpha_s^2}

\newcommand{\nlojet}{{\sc nlojet++}}

\newcommand{\asn}{\alpha_s^n}
\newcommand{\asd}{\alpha_s^d}

\newcommand{\sign}{\sigma_{\rm n}}
\newcommand{\sigd}{\sigma_{\rm d}}
\newcommand{\signi}[1]{\sigma_{n,#1}}
\newcommand{\sigdi}[1]{\sigma_{d,#1}}

\newcommand{\kf}[2]{{k_{#1,#2}}}

\title{Perturbative QCD predictions in fixed order for cross section ratios}

\author{L. Sawyer,}
\author{C. Waits,}
\author{and M. Wobisch}
\affiliation{Department of Physics, Louisiana Tech University,\\
             600 Dan Reneau Dr., Ruston, USA}

\emailAdd{sawyer@latech.edu}
\emailAdd{connor.h.waits-1@ou.edu}
\emailAdd{wobisch@latech.edu}

\abstract{
  In the standard approach, predictions
  of perturbative Quantum Chromodynamics
  for ratios of cross sections are computed as the ratio
  of fixed-order predictions for the numerator and the denominator.
  Beyond the lowest order in the perturbative expansion,
  the result does, however, not correspond to a fixed-order prediction
  for the ratio.
  This article describes how exact fixed-order results for ratios
  of arbitrary cross sections can be obtained.
  The general method for computations in any order of the perturbative
  expansion is derived, and results
  for next-to-leading order
  and next-to-next-to-leading order calculations are given.
  The approach is applied to theory predictions for various
  multi-jet cross section ratios measured at hadron colliders.
  The two methods are compared with each other and
  with the experimental data.
  Recommendations are made how to obtain improved
  theory predictions with more realistic uncertainty
  estimates.
}

\begin{document} 
\maketitle
\flushbottom

\section{Introduction}
\label{sec:intro}

Measurable quantities, defined as ratios of cross sections, $R$,
are studied for a variety of reasons.
Due to the cancellation of uncertainties which are correlated
between numerator and denominator,
such ratios can often be measured
with higher experimental precision
and also be more precisely predicted by theory.
Among such quantities are normalized distributions
where a differential cross section is divided by its integral,
or ratios of production cross sections for different processes.
Examples for the latter are ratios of multi-jet cross sections
in hadron collisions with different jet multiplicities,
which are frequently used for phenomenological tests
of perturbative Quantum Chromodynamics, pQCD,
for determinations of the strong coupling constant, $\as$,
and for tests of the renormalization group predictions for its
running~\cite{D0:2012xif,D0:2012dcq,D0:2012dqa,CMS:2013vbb,ATLAS:2018sjf}.

In most cases, phenomenological analyses of experimental data
use pQCD predictions in fixed order in $\as$.
The pQCD predictions for a ratio $R$ are then typically computed
from the ratio of the corresponding fixed-order predictions
for the cross sections in the numerator and the denominator.
For consistency, both of these are computed at the same
relative order in $\as$, i.e.\ both at leading order, LO,
next-to-leading order, NLO, or next-to-next-to-leading order, NNLO,
in $\as$.
Beyond the LO, however, such computations do not provide exact
fixed-order pQCD predictions for the ratio $R$.

In this article, we discuss how the
cross section results for the numerator and denominator
can be used to compute an exact fixed-order result for
the ratio $R$ of arbitrary cross sections.
We present the general procedure for any order pQCD,
and provide the solutions for predictions at NLO and NNLO.
Specific results for both methods are computed
for a set of multi-jet cross section ratios, measured at the
CERN LHC and the Fermilab Tevatron Collider.
From the comparison of both methods and their
description of the experimental data,
and based on general considerations,
we derive recommendations
for theoretical calculations of cross section ratios.
We motivate how this can lead to improved predictions
with more realistic uncertainty estimates
in future phenomenological analyses of cross section ratios.

\section{Calculation}
\label{sec:calc}

%
\subsection*{Definitions}

A measurable quantity $R = \frac{\sign}{\sigd}$
is defined as the ratio of two cross sections $\sign$ and $\sigd$,
where $n$ and $d$ denote the order of $\as$ of the LO contributions
to the pQCD predictions
($\signi{\rm LO} \propto \asn$ and $\sigdi{\rm LO} \propto \asd$)
and it is assumed that $n \ge d$.
It is also assumed that the
quantity $R$ is defined in bins of an energy
or transverse momentum related variable $p$
which is defined for both $\sign$ and $\sigd$.
At a fixed value (or in a given bin) of $p$,
the quantity $R$ is given by 
$R(p) = \frac{\sign(p)}{\sigd(p)}$,
and it is assumed that in a pQCD calculation
the renormalization scale $\mur$
can be related to $p$ by the same simple function
(like $\mur = p$ or $\mur = p/2$) for both, $\sign$ and $\sigd$.
In other words, in a given bin of $p$, the ratio $R$ is probing
$\as$ and the pQCD matrix elements for $\sign$ and $\sigd$
at the same $\mur$.
In the following, for the sake of brevity, the dependence on $p$ is
omitted
The perturbative expansions of the two cross sections
in orders of $\as$ are given by
\begin{equation}
  \sign = \sum_{i=0}^{\infty} \, \signi{i}
  \qquad  \hbox{and} \qquad 
  \sigd = \sum_{i=0}^{\infty} \, \sigdi{i} \, ,
\end{equation}
where the pQCD contribution in LO corresponds to
$i=0$, NLO to $i=1$, and NNLO to $i=2$.

Typically, the ``NLO $k$-factor'', $k_{\rm NLO}$,
for a given quantity is defined
as the ratio of its NLO and LO pQCD predictions.
For $\sign$ it is therefore
\begin{equation}
  k_{\rm NLO} =  \frac{\signi{\rm NLO}}{\signi{\rm LO}} \, = \,
  \frac{\signi{0}+\signi{1}}{\signi{0}} \, = \,
  1 + \frac{\signi{1}}{\signi{0}}.
\end{equation}
We name the last piece $ \kf{n}{1} \equiv \frac{\signi{1}}{\signi{0}} $
a ``reduced NLO $k$-factor'', since  $\kf{n}{1} =  (k_{\rm NLO} - 1$).
This definition is then extended to all orders, such that
for each order $i$, a reduced $k$-factor is defined as
\begin{equation}
  \kf{n}{i} = \frac{\signi{i}}{\signi{0}}
  \quad \mbox{for} \quad i = 1,2,3,\cdots  \, .
  \label{eq:kfac}
\end{equation}
With this definition, the perturbative expansions of $\sign$ and $\sigd$
can be written
in terms of their LO contributions and their reduced $k$-factors as
\begin{equation}
  \sign = \signi{0} \cdot (1 + k_{n,1} + k_{n,2} + ...)
  \quad \mbox{and} \quad
  \sigd = \sigdi{0} \cdot (1 + k_{d,1} + k_{d,2} + ...)   \, .
  \label{eq:sigmaexpansion}
\end{equation}

%
\subsection*{The standard approach for computing the ratio}

The LO pQCD prediction for the ratio $R = \frac{\sign}{\sigd}$
is uniquely given by
the ratio of the LO pQCD results for the numerator and denominator
\begin{equation}
  R_{\rm LO} = \frac{\signi{\rm LO}}{\sigdi{\rm LO}}
  = \frac{\signi{0}}{\sigdi{0}}    \, .
  \label{eq:rlo}
\end{equation}
Beyond LO, however, the pQCD prediction for $R$
can be obtained in different ways.
In phenomenological analyses of experimental data,
the NLO (NNLO) pQCD prediction for $R$ is usually computed 
from the ratio of the NLO (NNLO) predictions for $\sign$ and $\sigd$ as
\begin{equation}
  R_{\rm NLO} = \frac{\signi{\rm NLO}}{\sigdi{\rm NLO}}
  \qquad   \mbox{and} \qquad
  R_{\rm NNLO} = \frac{\signi{\rm NNLO}}{\sigdi{\rm NNLO}}   \, .
  \label{eq:standard}
\end{equation}
With \eqref{eq:sigmaexpansion} and \eqref{eq:rlo},
the NLO result can be written as
\begin{equation}
  R_{\rm NLO} = \frac{\signi{0}\cdot(1+\kf{n}{1})}{\sigdi{0}\cdot(1+\kf{d}{1})}
  \, = \,   R_{\rm LO} \cdot \frac{1+\kf{n}{1}}{1+\kf{d}{1}}  \, ,
  \label{eq:standardnlo}
\end{equation}
the NNLO result as
\begin{equation}
  R_{\rm NNLO} = \frac{\signi{0}\cdot(1+\kf{n}{1}+\kf{n}{2})}{\sigdi{0}\cdot(1+\kf{d}{1}+\kf{d}{2})}
  \, = \,
   R_{\rm LO} \cdot \frac{1+\kf{n}{1}+\kf{n}{2}}{1+\kf{d}{1}+\kf{d}{2}} \, ,
    \label{eq:standardnnlo}  
\end{equation}
and the general expression as
\begin{equation}
  R_{\rm fixed\mbox{-}order} = \frac{\signi{0}\cdot(1+\sum_{i=1}^{i_{\rm max}}
    \kf{n}{i})}{\sigdi{0}\cdot(1+\sum_{i=1}^{i_{\rm max}} \kf{d}{i})}  
  \, = \,   R_{\rm LO} \cdot
  \frac{1+\sum_{i=1}^{i_{\rm max}} \kf{n}{i}}{1+\sum_{i=1}^{i_{\rm max}} \kf{d}{i}}
  \, ,
  \label{eq:standardall}
\end{equation}
where $i_{\rm max}$ specifies the highest order in the pQCD calculations
(1: NLO, 2: NNLO, ...).
In the following, we refer to the results
\eqref{eq:standard}--\eqref{eq:standardall},
as the ``standard'' approach.
Note that, beyond LO, these ratios do not directly correspond to
fixed-order pQCD results for the quantity $R$.

%
\subsection*{Computing the ratio at fixed order}

The goal is to obtain an exact fixed-order result for $R$, in the form
\begin{equation}
    R = R_{\rm LO} \cdot \left( 1 + \sum_{i=1}^{i_{\rm max}} \as^i c_i \right)  \, .
\end{equation}
To find the terms of the sum, we rewrite \eqref{eq:standardall} as
\begin{equation}
  R_{\rm fixed\mbox{-}order} = R_{\rm LO} \cdot \left(1+\sum_{i=1}^{i_{\rm max}} \kf{n}{i}\right)
  \cdot \frac{1}{1+\sum_{i=1}^{i_{\rm max}} \kf{d}{i}}    \, ,
  \label{eq:startingpoint}
\end{equation}
and expand the right term in a Taylor series
\begin{equation}
  \frac{1}{1+x} = 1 - x +x^2 - x^3 + x^4 - \cdots
  \hskip4mm \mbox{with} \hskip4mm
  x = \sum_{i=1}^{i_{\rm max}} \kf{d}{i}  \; .
\end{equation}
The terms of this series are multiplied with the terms
in parenthesis in \eqref{eq:startingpoint},
and the resulting products are sorted in their powers of $\as$.
The infinite series is then truncated at the corresponding order
at which $\sign$ and $\sigd$ were
computed.\footnote{The sorting is facilitated by the fact
that the $\kf{n}{i}$ and $\kf{d}{i}$ are both proportional to $\as^i$,
which follows directly from their definition in \eqref{eq:kfac}.}
This can be done at any order pQCD.
To compute a corresponding fixed-order pQCD result for $R$ requires,
of course, to compute both $\sign$ and $\sigd$ at the same corresponding
relative order in $\as$ (e.g.\ NLO, NNLO, ...).

%
\subsection*{Results for NLO and NNLO}

At NLO, with $x = \kf{d}{1}$ in the Taylor series, one obtains
\begin{equation}
  R_{\rm NLO} = R_{\rm LO} \cdot (1 + \kf{n}{1} - \kf{d}{1} )  \, ,
  \label{eq:fixednlo}
\end{equation}
and at NNLO, with $x = (\kf{d}{1}+\kf{d}{2})$ in the Taylor series,
\begin{equation}
  R_{\rm NNLO} = R_{\rm LO} \cdot \left[ 1 + (\kf{n}{1} - \kf{d}{1})
    + (\kf{n}{2} - \kf{d}{2}) -  \kf{d}{1} (\kf{n}{1}- \kf{d}{1})
    \right]   \, .
  \label{eq:fixednnlo}
\end{equation}
The individual pieces in \eqref{eq:fixednlo} and \eqref{eq:fixednnlo}
can simply be computed from the LO and (N)NLO pQCD results
from the cross section calculations for $\sign$ and $\sigd$.
The calculation of $R_{\rm NLO}$ requires
\begin{equation}
  R_{\rm LO} = \frac{\signi{\rm LO}}{\sigdi{\rm LO}} \; ,
  \qquad
  \kf{n}{1} = \frac{\signi{\rm NLO} - \signi{\rm LO}}{\signi{\rm LO}} \; ,
  \qquad
  \kf{d}{1} = \frac{\sigdi{\rm NLO} - \sigdi{\rm LO}}{\signi{\rm LO}} \; ,
\end{equation}
and the calculation of $R_{\rm NNLO}$ requires in addition
\begin{equation}
  \kf{n}{2} = \frac{\signi{\rm NNLO} - \signi{\rm NLO}}{\signi{\rm LO}} \, ,
  \qquad
  \kf{d}{2} = \frac{\sigdi{\rm NNLO} - \sigdi{\rm NLO}}{\signi{\rm LO}} \, .
\end{equation}

%
\subsection*{The difference of the two results at NLO}

The fixed-order result \eqref{eq:fixednlo} differs from the 
``standard'' result \eqref{eq:standardnlo}
due to the truncation of the series of the products
of the terms in parenthesis in \eqref{eq:startingpoint}
and the terms from the Taylor series.
An estimate of this difference
can be obtained from the leading term of the truncated piece.
At NLO, this is the term proportional to $\ass$.
The difference of the results from the two methods is 
\begin{equation}
  R_{\rm NLO, fixed\mbox{-}order} - R_{\rm NLO, standard} \, = \, R_{\rm LO}  \cdot
  \left[
    \kf{d}{1} \cdot (\kf{n}{1} - \kf{d}{1}) \right] \, + \,
  \mbox{higher orders}
  \, .
  \label{eq:difference}
\end{equation}
This difference is proportional to the LO result,
$R_{\rm LO}$, and it depends on the NLO $k$-factors
for the numerator and denominator.
It becomes small if either
the NLO $k$-factor for the denominator is close to unity
(corresponding to a reduced $k$-factor of $|\kf{d}{1}| \ll 1$)
and/or if the NLO $k$-factors for the numerator and denominator
become equal ($\kf{n}{1} \simeq \kf{d}{1}$).
It is interesting to note the asymmetry:
For the difference to become small, it is sufficient
that the $k$-factor for the denominator is small
while it is not affected by the $k$-factor for the numerator alone.
Based on \eqref{eq:difference}, the two $k$-factors can be used
to obtain a quick estimate of the difference between
the ``standard'' method and the fixed-order result at NLO.

\section{Comparison with data}
\label{sec:data}

\begin{figure}  
\centering
\includegraphics[width=.53\textwidth]{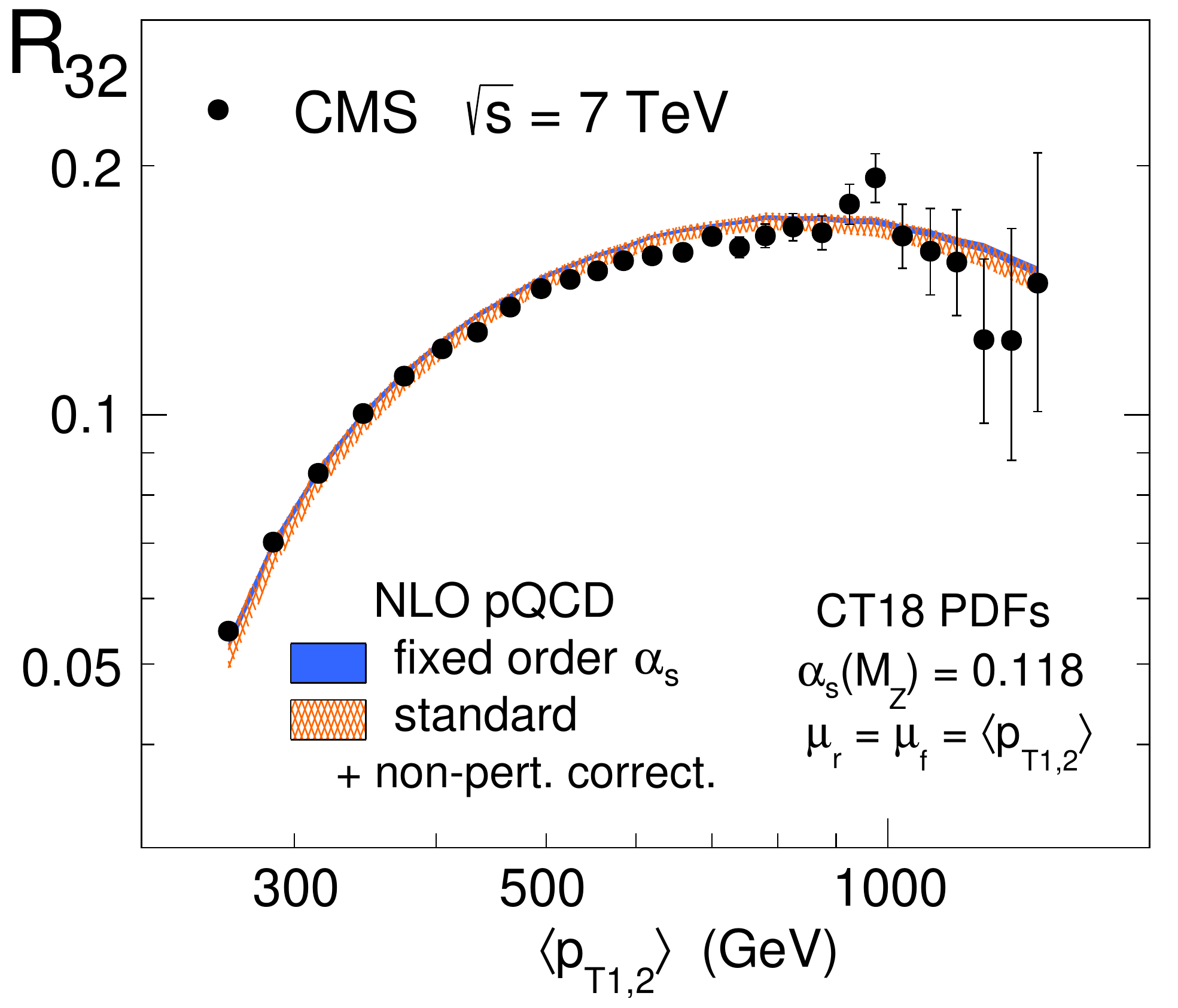} 
\caption{\label{fig:cmsr32}
  The multi-jet cross section ratio $\Rtt$, measured in $pp$ collisions
  at $\sqrt{s}=7$\,TeV  as a function of $\langle p_{T1,2} \rangle$
  in the CMS experiment~\cite{CMS:2013vbb}.
  Two sets of pQCD predictions,
  corrected for non-perturbative contributions,
  are compared to the data: 
  the fixed-order results for $\Rtt$, and the results from the
  ``standard'' approach, computed from the ratio of the fixed-order
  results for the two cross sections.
  The shaded areas represent the ranges of the scale dependences
  of the calculations.
}
\end{figure}

In the following, we use the formulae obtained in the previous section
to compute NLO pQCD predictions for selected quantities
which are then compared to the results from experimental measurements.
For this purpose, we focus on five measurements of different multi-jet
cross section ratios at the CERN LHC (in $pp$ collisions at
$\sqrt{s} = 7\,$TeV and 8\,TeV) and the Fermilab Tevatron Collider
(in $p\bar{p}$ collisions at $\sqrt{s} = 1.96$\,TeV).
These include measurements of the quantities  $\Rtt$, $\Rdphi$, and $\Rdr$,
which are different ratios of three-jet and two-jet production
processes.

The theoretical predictions for the ratios at NLO
are obtained from the
LO and NLO pQCD results for the two-jet and three-jet cross section
calculations which are computed using \nlojet~\cite{Nagy:2001fj,Nagy:2003tz}
with fastNLO~\cite{Kluge:2006xs,Britzger:2012bs}.
Parametrizations of the parton distribution functions, PDFs, of
the proton are taken from the results of the
global analysis CT18~\cite{Hou:2019efy}.
The renormalization, $\mur$, and factorization scales, $\muf$,
are set to the same values
as used in the experimental publications of the measurement results,
either to one of the relevant jet $p_T$ variables, or to half of the
total jet $p_T$ sum, $H_T/2$.
The uncertainty of the pQCD results due to the $\murf$ dependence
is computed from independent variations of $\mur$ and $\muf$
by factors of 0.5--2 around the nominal choices.
The corresponding range of variations is referred to as ``scale dependence''.
Correction factors, to account for non-perturbative contributions
are taken from
the estimates that were obtained in the experimental analyses.
PDF uncertainties are not relevant for the following discussions
and have not been evaluated.
Computations according to \eqref{eq:fixednlo} are referred to
as ``fixed-order'' results,
and those, based on the left equation in \eqref{eq:standard}
as the ``standard'' method.

\begin{figure}  
\centering
\includegraphics[width=0.99\textwidth]{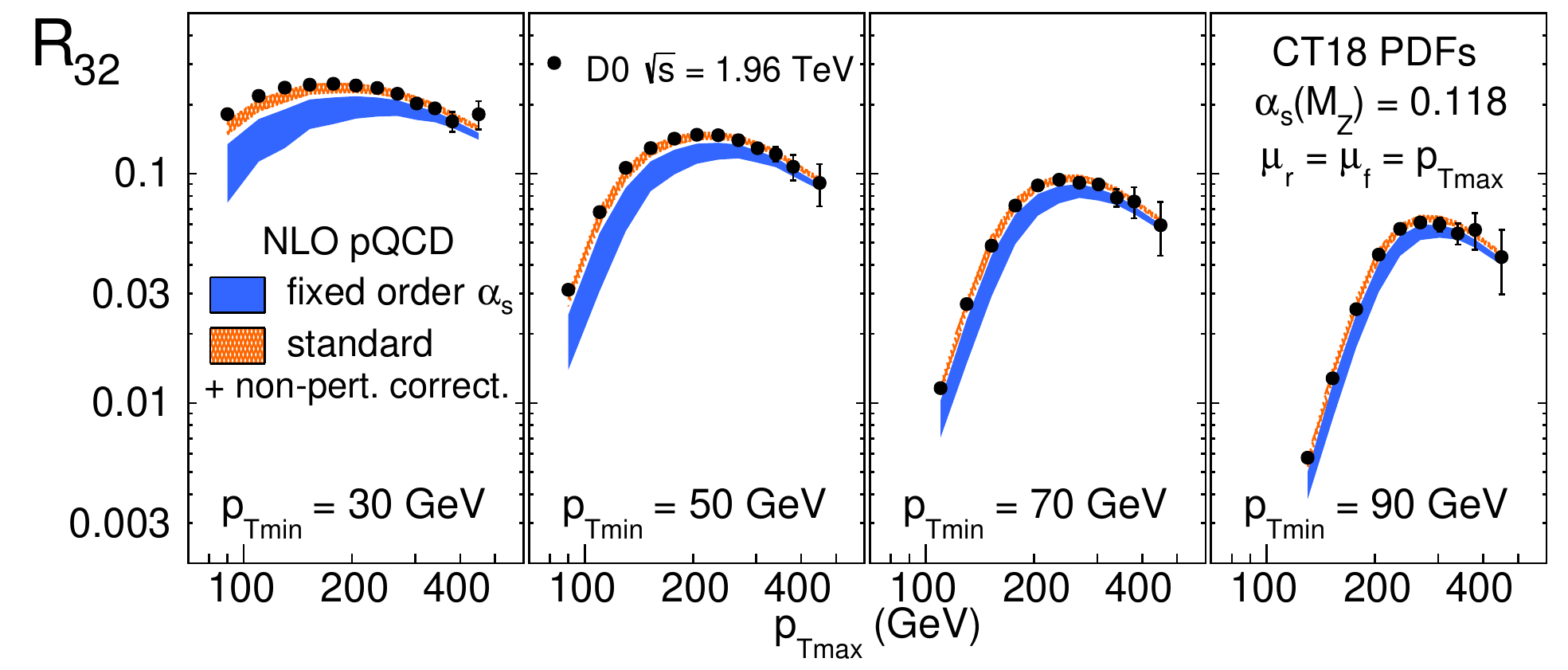}
\caption{\label{fig:d0r32}
  The multi-jet cross section ratio $\Rtt$, measured in $p\bar{p}$ collisions
  at $\sqrt{s}=1.96$\,TeV  as a function of $\ptmax$
  in the D0 experiment~\cite{D0:2012dcq}.
  Two sets of pQCD predictions,
  corrected for non-perturbative contributions,
  are compared to the data: 
  the fixed-order results for $\Rtt$, and the results from the
  ``standard'' approach, computed from the ratio of the fixed-order
  results for the two cross sections.
  The shaded areas represent the ranges of the scale dependences
  of the calculations.
}
\end{figure}

%
The CMS Collaboration has measured the ratio
of the inclusive three-jet cross section and
the inclusive two-jet cross section, $\Rtt$,
for jets with $p_T > 150\,$GeV and rapidities of $|y|<2.5$~\cite{CMS:2013vbb}.
The results are published as a function of the
average transverse momentum
of the two leading jets in the event,
$\langle p_{T1,2} \rangle$,
over the range
$0.42 < \langle p_{T1,2} \rangle <1.39$\,TeV,
as displayed in Figure~\ref{fig:cmsr32}. 
The results of the fixed-order and ``standard'' calculations
for $\Rtt$ are compared to the data, with their error bands
representing the range of their respective scale dependences.
Both results are in agreement, and both describe the data
equally well.

%
Another measurement of the ratio $\Rtt$
was made by the D0 Collaboration
for jets with rapidities $|y|<2.4$
and for various lower jet $p_T$ requirements, $\ptmin$.
The results are published as a function
of the leading jet $p_T$, $\ptmax$, 
over the range $80 < \ptmax < 500\,$GeV
and for four $\ptmin$ choices from 30 to 90\,GeV,
as displayed in Figure~\ref{fig:d0r32}.
The pQCD prediction from the ``standard'' method gives
a reasonable description of the data over the whole $\ptmax$ range
for all $\ptmin$ requirements, except for the lowest $\ptmax$ bins
for $\ptmin = 30\,$GeV.
The fixed-order calculation predicts lower values everywhere
and the scale uncertainty bands of the two calculations do either
not, or hardly, overlap.
Only towards larger $\ptmin$ and larger $\ptmax$, the uncertainty bands
get closer.

\begin{figure}  
\centering
\includegraphics[width=.85\textwidth]{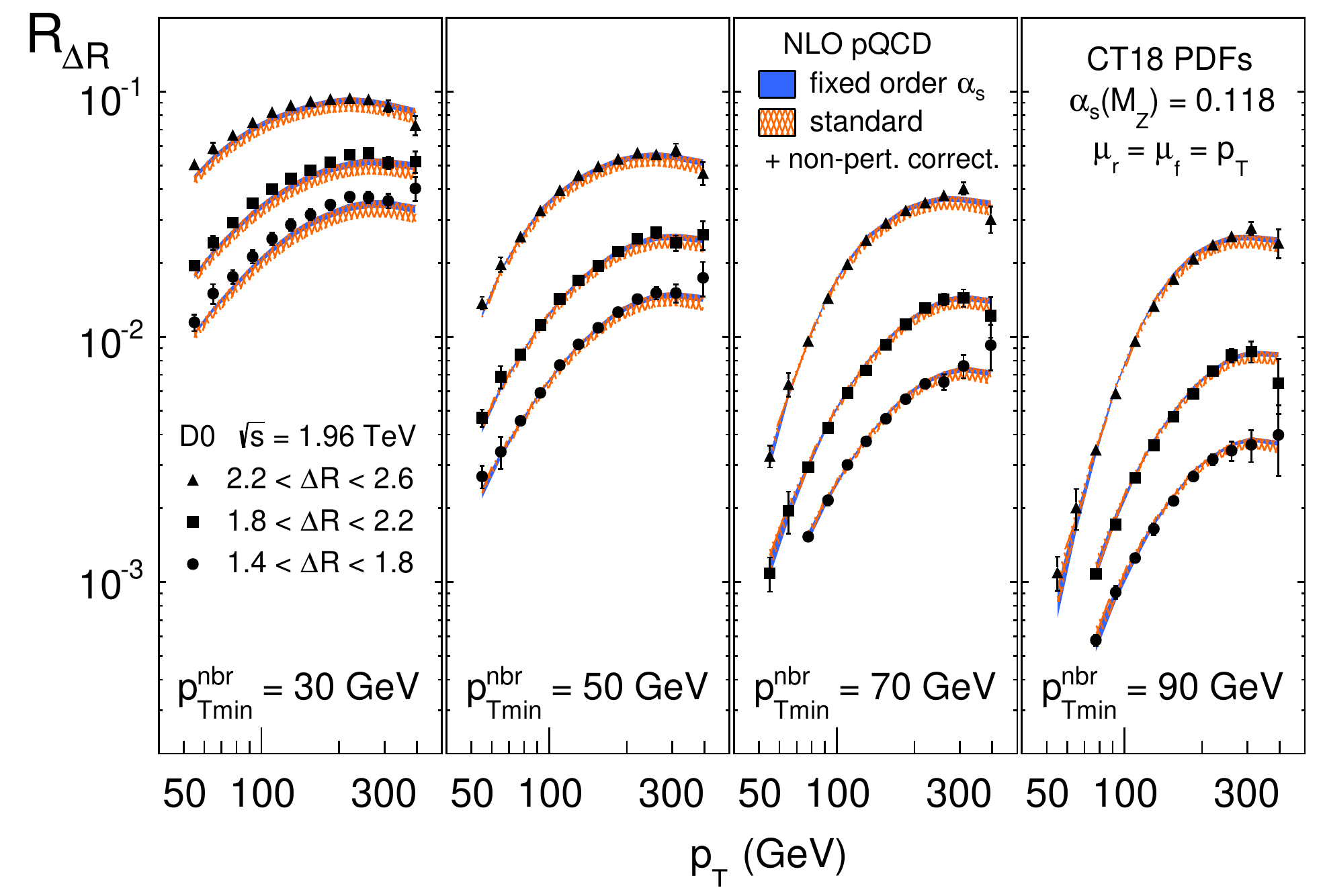} 
\caption{\label{fig:d0rdr}
  The multi-jet cross section ratio $\Rdr$, measured
  in $p\bar{p}$ collisions at $\sqrt{s}=1.96$\,TeV
  in the D0 experiment~\cite{D0:2012xif},
  as a function of $p_T$, in four values of $\ptngbmin$ and
  in three regions of $\DR$.
  Two sets of pQCD predictions,
  corrected for non-perturbative contributions,
  are compared to the data:
  the fixed-order results for $\Rdr$, and the results from the
  ``standard'' approach, computed from the ratio of the fixed-order
  results for the two cross sections.
  The shaded areas represent the ranges of the scale dependences
  of the calculations.
}
\end{figure}

\begin{figure}  
\centering
\includegraphics[width=.92\textwidth]{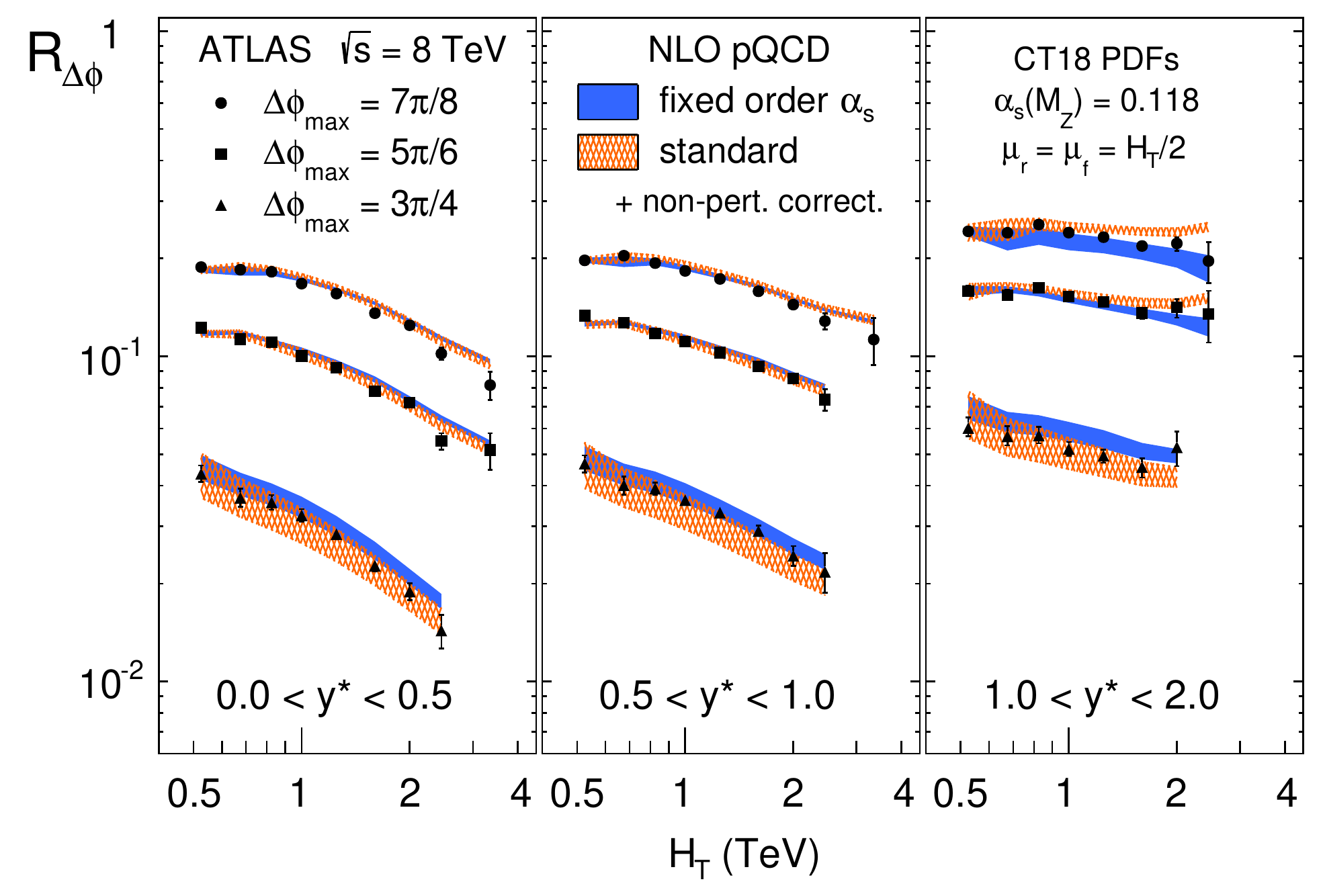}
\includegraphics[width=.92\textwidth]{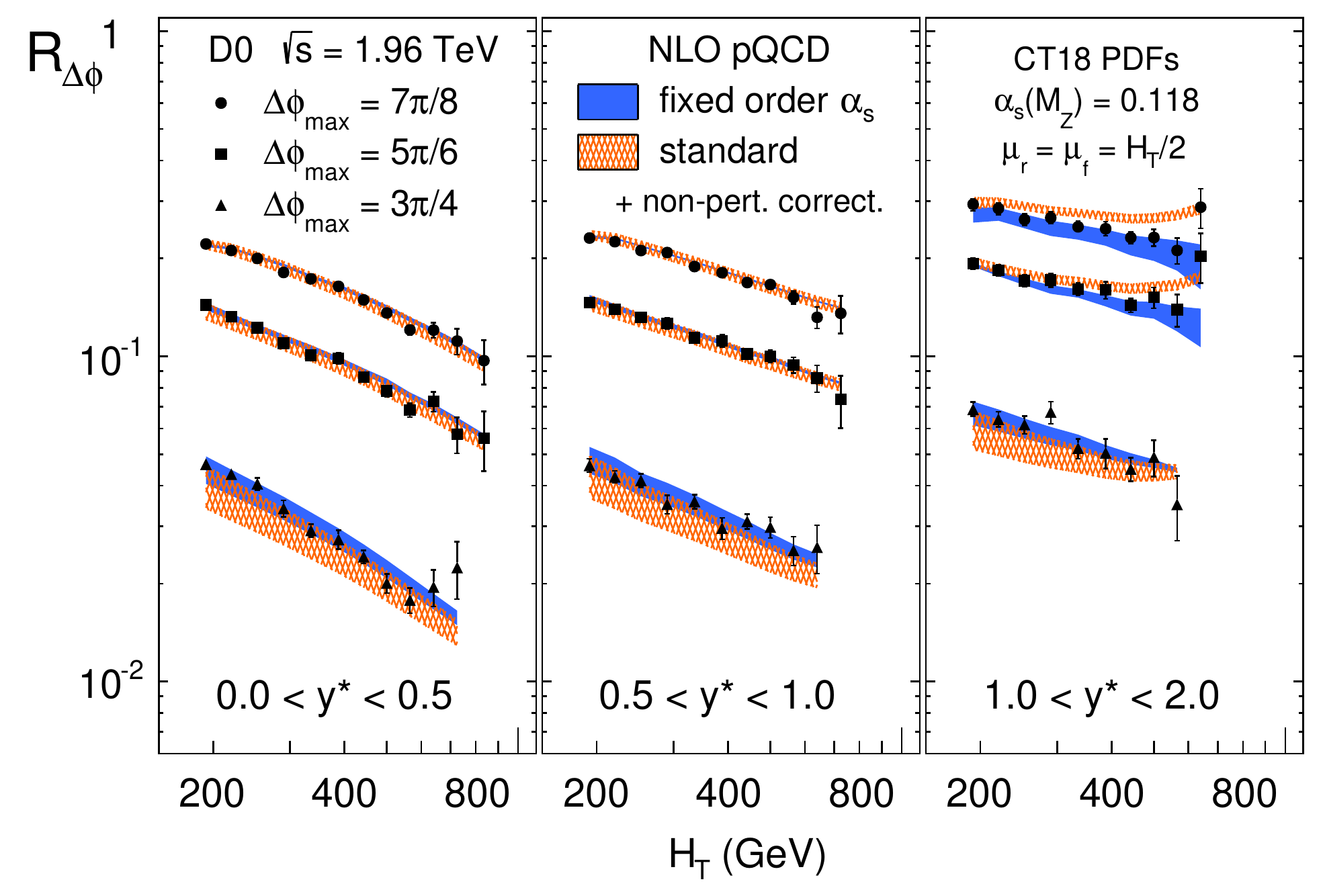}
\caption{\label{fig:atlrdphi}
  The multi-jet cross section ratio $\Rdphi$, measured
  in $pp$ collisions at $\sqrt{s}=8$\,TeV
  in the ATLAS experiment~\cite{ATLAS:2018sjf} (top)
  and in $p\bar{p}$ collisions at $\sqrt{s}=1.96$\,TeV
  in the D0 experiment~\cite{D0:2012dqa} (bottom),
  as a function of $H_T$, in three regions of $y^*$ and for
  three values of $\Dphimax$.  
  Two sets of pQCD predictions,
  corrected for non-perturbative contributions,
  are compared to the data:
  the fixed-order results for $\Rdphi$, and the results from the
  ``standard'' approach, computed from the ratio of the fixed-order
  results for the two cross sections.
  The shaded areas represent the ranges of the scale dependences
  of the calculations.
}
\end{figure}

%
The D0 Collaboration also published measurement results
of a new quantity, $\Rdr$~\cite{D0:2012xif}.
This quantity, while defined in a more inclusive way,
also probes the ratio of three-jet and two-jet production.
Starting point is an inclusive jet sample (which probes the
two-jet production process).
The presence of a neighboring jet with $\Delta R < \pi$
is a sign of an event topology with three or more jets.
The fraction of all inclusive jets with a neighboring jet, $\Rdr$,
is therefore also a three- over two-jet ratio.
The quantity $\Rdr$ was measured
for different $p_T$ requirements, $\ptngbmin$,
and different angular separations, $\DR$, for the neighboring jets,
as a function of inclusive jet $p_T$ from
50 to 450\,GeV.
The results of the fixed-order and the ``standard'' calculations
for $\Rdr$ are compared to the data in Figure~\ref{fig:d0rdr}.
In almost all of the phase space the conclusions mirror
those for the theoretical description of the CMS $\Rtt$ data in
Figure~\ref{fig:cmsr32}:
The fixed-order pQCD predictions agree with those from
the ``standard'' method,
and both give a good description of all data with $\ptngbmin \ge 50\,$GeV.
Only in the softer regime, for $\ptngbmin = 30\,$GeV at smaller $p_T$,
they slightly underestimate the experimental measurement results.

\pagebreak

%
The multi-jet ratio $\Rdphi$ was proposed~\cite{Wobisch:2012au}
as another ratio of three- and two-jet production processes
that probes the azimuthal decorrelations of the two leading $p_T$ jets
in an event.
The $\Rdphi$ measurements by the D0 and ATLAS
Collaborations~\cite{D0:2012dqa,ATLAS:2018sjf}
both follow the recommendations from the original proposal,
and the two results are shown in Figure~\ref{fig:atlrdphi}.
Both measurements are performed in the same three
rapidity regions, $y^*$,
and for the same azimuthal decorrelation requirements, $\Dphimax$.
The ATLAS (D0) data are presented as a function of
the scalar  $p_T$ sum of all jets in an event, $H_T$,
over the range  0.46--4\,TeV  (180--900\,GeV).
Since both data sets are measured at different center-of-mass energies,
$\sqrt{s}$, their results should be compared to each other
at the same $H_T/\sqrt{s}$.
The degree of agreement between the NLO pQCD predictions
from the fixed-order and the ``standard'' calculations
and how they describe the data is pretty much the same
for the ATLAS and D0 data sets.
In different  $y^*$ and $\Dphimax$ regions, however,
the two calculations exhibit a rather different behavior.
At $y^* < 1.0$, for $\Dphimax = 7\pi/8$ and $5\pi/6$,
both calculations agree very well,
exhibit a relatively small scale dependence,
and both describe the data.
At $\Dphimax = 3\pi/4$,
the two predictions start to deviate from each other,
and in some cases
their larger scale uncertainty bands have only a small overlap.
The data are described by both predictions.
At $1 < y^* < 2$, for $\Dphimax = 7\pi/8$ and $5\pi/6$,
the two predictions have a different $H_T$ dependence
and disagree at high $H_T$.
In these regions,
the fixed-order calculation gives a better description
of the overall $H_T$ shape for both data sets.

%
\section{Summary and recommendations}

In the ``standard'' approach, pQCD predictions
for cross section ratios are computed
as the ratio of the fixed-order results of the numerator
and the denominator.
Such a calculation does, however, not represent an
exact fixed-order result for the ratio.
A method, how to obtain exact fixed-order results
for ratios of cross sections is presented in this article.
NLO pQCD predictions according to the new fixed-order method
and the ``standard'' method
were calculated for
five different experimental measurements
of multi-jet cross section ratios.
For these measured quantities the two predictions were compared
to each other and to the experimental data.
It was found that in all cases where the results from the two methods
agree with each other
(as seen for the CMS $\Rtt$, the D0 $\Rdr$,
and some regions of the ATLAS and D0 $\Rdphi$ measurements
in Figures~\ref{fig:cmsr32}, \ref{fig:d0rdr}, and \ref{fig:atlrdphi}),
they also both describe the data.
In all cases where the two methods disagree
(meaning that their scale uncertainty bands do not overlap,
as seen for the D0 $\Rtt$ data and the high $H_T$ tails
in some of the ATLAS and D0 $\Rdphi$ data at $y^* > 1$
in Figures~\ref{fig:d0rdr} and \ref{fig:atlrdphi}),
one of them (but not always the same) is describing the data.
In some intermediate cases, where the scale uncertainty bands from the
two methods have little overlap
(as for the ATLAS and D0 $\Rdphi$ data with $\Dphimax = 3\pi/4$
in Figure \ref{fig:atlrdphi}),
both predictions are somehow consistent with the data.

It is important here to note, that neither of the two methods
is fundamentally preferable over the other
and both stand exactly on the same footing.
In any given order pQCD, the results from both methods are
equally valid representation of the perturbative expansion,
and they only differ in
higher-order terms, just like for the scale dependence
of the calculation.\footnote{The results
of fixed-order pQCD calculations obtained at different scales $\murf$
differ only due to terms which are of higher orders of $\as$.
Since the scale dependence is a reflection of some
higher-order terms, it is typically used as an estimate
of the potential size of the uncalculated higher order
contributions.
}
Consequently, both results should be taken into account,
and their discrepancy 
should be regarded as a genuine uncertainty,
which is not always covered by the range
of the scale uncertainties of the individual methods
(for example in Figure~\ref{fig:atlrdphi}).

For future phenomenological studies of cross section ratios,
we make the following recommendations:
\begin{enumerate}
\item For the central pQCD prediction\\
  Typically, the goal of phenomenological studies is to find {\em if}
  a theory can describe the data. Therefore we recommend to consider
  both methods, and use the one that gives a better description
  of the data. In general, this can be decided by a $\chi^2$
  calculation (especially in the case of parameter determinations)
  or sometimes simply by eye (as in Figure~\ref{fig:d0r32}).

\item For the uncertainty of the prediction\\
  Based on the arguments given above, we recommend to use the
  full envelope of the scale uncertainty bands from both methods
  as an estimate of the pQCD uncertainty related to uncalculated
  terms of higher order in $\as$.
\end{enumerate}

While sometimes the two methods differ only in their normalization
(as in Figure~\ref{fig:d0r32}), in other cases they can also
feature very different shapes (as in Figure~\ref{fig:atlrdphi}).
The first recommendation ensures that one does not rule out a
theory which, in an equally valid alternative form,
would actually be able to describe a given data set.
The second recommendation acknowledges the additional
uncertainty contribution related to the difference of
two equally valid representations of the pQCD calculation
which may exceed the uncertainty estimate based on simple scale variations
(as seen in Figure~\ref{fig:atlrdphi}).
This will provide more realistic estimates of theoretical uncertainties
and will be helpful in future phenomenological studies
when identifying robust measurable quantities
for which the theoretical approximations are more reliable.

\acknowledgments

This work is supported by grant 1913877
from the National Science Foundation.
L.S.\ and M.W.\ also wish to thank the Louisiana Board of Regents
Support Fund for the support through the Charles \& Nelwyn Spruell 
and the Eva J. Cunningham Endowed Professorships.

\bibliography{pqcdratio}

\end{document}